\documentclass[letterpaper, 10 pt, conference]{ieeeconf}  % Comment this line out if you need a4paper
\IEEEoverridecommandlockouts
\usepackage{cite}
\usepackage{amsmath,amssymb,amsfonts}
\usepackage{algorithmic}
\usepackage{graphicx}
\usepackage[hyphens,spaces,obeyspaces]{url}
\usepackage{hyperref}
\usepackage{todonotes}
\usepackage{textcomp}
\usepackage{xcolor}
\usepackage{bm}
\usepackage{threeparttable}
\usepackage{siunitx}
\usepackage{multirow}
\usepackage{subfig}
\usepackage{comment}
\usepackage{tikz}
\usepackage{url}
\usepackage{algorithm2e}
\usepackage{caption}
\usepackage{hyperref}
\usepackage{lipsum}
\usepackage{gensymb}
\usepackage{booktabs}
\usetikzlibrary{shapes}

\usepackage[skip=8pt,font=small]{caption}

\def\BibTeX{{\rm B\kern-.05em{\sc i\kern-.025em b}\kern-.08em
    T\kern-.1667em\lower.7ex\hbox{E}\kern-.125emX}}

\hypersetup{
	colorlinks=true,
	linkcolor=black,
	filecolor=magenta,      
	urlcolor=black,
}

\begin{document}

% \title{\LARGE \bf Comparing Human perception of a mobile robots driving behavior in a virtual and a real environment}

\title{\LARGE \bf Perceived Safety of Workers in Encounters with Large Industrial AGVs}
%Perceived Safety of Workers in Encounters with Large Industrial AGVs
%Real or Virtual? Comparing Worker Perception of an Industrial AGV's Driving Behavior Across Modalities

\author{Ansgar~Howey$^{1}$,
       Tim~Schreiter$^{2}$,
       Andrey Rudenko$^{2,3}$ 
       and~Achim J.~Lilienthal$^{2,3,4}$% <-this % stops a %space
\thanks{$^{1}$KION Supply Chain Solutions, Still GmbH, Hamburg, Germany
{\tt\small ansgar.howey@kiongroup.com}}%
\thanks{$^{2}$Chair Perception for Intelligent Systems, TUM, Germany 
{\tt\small \{tim.schreiter,andrey.rudenko\}@tum.de}}%
\thanks{$^{3}$Munich Institute of Robotics and Machine Intelligence (MIRMI), TUM, Germany~~{\tt\small achim.j.lilienthal@tum.de}}%
\thanks{$^{4}$Centre for Applied Autonomous Sensor Systems (AASS),
	\"Orebro University, Sweden}
    \thanks{This work has been supported by the German Federal Ministry of Research, Technology and Space (BMFTR) under the Robotics Institute Germany (RIG)}
    \thanks{This research was supported within the project IMOCO4.E which received funding from the Electronic Component Systems for European Leadership Joint Undertaking, under Grant Agreement n° 101007311}}

% Source - https://tex.stackexchange.com/a/745153
% Posted by Michael Dorner
% Retrieved 2026-05-11, License - CC BY-SA 4.0

% \author{Anonymous Authors}

\maketitle

\begin{abstract}
%Recently, Automatic Guided Vehicles (AGVs) are increasingly equipped with autonomous capabilities, which makes it important to evaluate their effects on humans encountering them. Since User Experience (UX) Testing in a simulation using virtual reality (VR) would significantly reduce risk and effort a comparison of a real-world and a virtual setting using a simple traffic scenario was conducted. Test persons recruited from the workforce of an industrial assembly line and the internal transport unit (forklift trucks) encountered an industrial AGV based on a series of trucks with a 1.200kg load capacity in reality and in VR. The perceived distances and speed, as well as the perceived threat level, were generally comparable with a slight offset. An UX evaluation using a VR setup seems feasible, though with some limitations.

Automated Guided Vehicles (AGV) in factory automation are increasingly capable of moving autonomously in close proximity to human workers.
%To enable seamless integration into such environments, these vehicles should exhibit high levels of both physical and perceived safety.
While their physical safety is regulated by standards and directives, perceived safety and workers' comfort in close-proximity interactions are being actively investigated in studies. There are three limitations in the prior art research to that end. Firstly, AGVs with larger payloads are understudied. Secondly, the test participants are usually students and not working professionals. Thirdly, while conducting in-person experiments with heavy machinery can be dangerous, the transfer of safety perception results from simulated experiments remains open. In this paper, we investigate industrial workers' perceived safety in shared spaces with large AGVs in a real-world encounter and in virtual reality. We vary the passing distance and the shape of the collision avoidance maneuver, and evaluate perceived threat level using a handheld pressure-sensitive trigger interface and a post-experiment questionnaire. Additionally, we ask participants to set their own collision avoidance parameters based on their experience with the demonstrated trajectory profiles. In a within-subject study, we found that, while the threat levels are perceived overall slightly higher in VR, the passing distance of 1.5 to 2 meters is preferred among the demonstrated profiles, as well as in the self-defined trajectories.

\end{abstract}

\section{Introduction}
%[HA1.1]
The use of Automated Guided Vehicles (AGVs) in industrial settings and warehouses has increased significantly in recent years \cite{haney2024literature}. The majority of industrial AGVs have a load capacity of 1-8 tons, maximum operating speed of 2-3 m/s, and their \textit{physical} safety is regulated by the European Machine Directive \cite{EU:Machinery:2006} and ISO 3691-4 \cite{ISO3691-4}.
%The most relevant harmonized standard is ISO 3691-4. To comply with this standard, AGVs are equipped with personal safety devices that observe the area in the driving direction. If any object is detected, a safety stop is triggered. 
AGVs usually operate in mixed traffic, including pedestrians and manually driven industrial trucks, and can be categorized into autonomous and non-autonomous trucks \cite{haney2024literature, han2025interactions}. AGVs without autonomous functions strictly follow predefined trajectories, making their paths predictable to people in regular contact with them \cite{howey2023user}.
%In order to enable them to react on current traffic situations or obstacles in their path or even freely plan their trajectories, AGVs have increasingly been equipped with autonomous \cite{scherb2023design} (???) capabilities, including the ability to understand their surroundings, plan routes autonomously, and identify and locate humans. In the following, these AGVs shall be referred to as Mobile Robots (MR).
With the increase in autonomous route planning and collision avoidance, their trajectories become less predictable.
%[TJ3.1][HA3.2][HA3.3]
This makes it important to evaluate their effect on the humans encountering them and draw design implications from these encounters, thus improving their legibility and achieving higher levels of \textit{perceived} safety. User Experience (UX) tests \cite{howey2023user} have shown that both explicit visual and auditory signals, as well as implicit information conveyed through the vehicle's driving behavior, can make its movement more predictable. Prior work has dealt with explicit external signals and lead to the VDMA recommendation ``Design of visible and audible signals of driverless industrial trucks" \cite{scherb2023design}. At the same time, there are indications that people actually rely mainly on the robot's current movement to anticipate its future path and do not consider visual or auditory signals at all \cite{howey2023user}. This motivates the movement of the robot as a potential factor in legibility and perceived safety.

\begin{figure}[!t]
    \centering
    \subfloat[][Real-world encounter]{%
        \includegraphics[width=0.48\columnwidth]{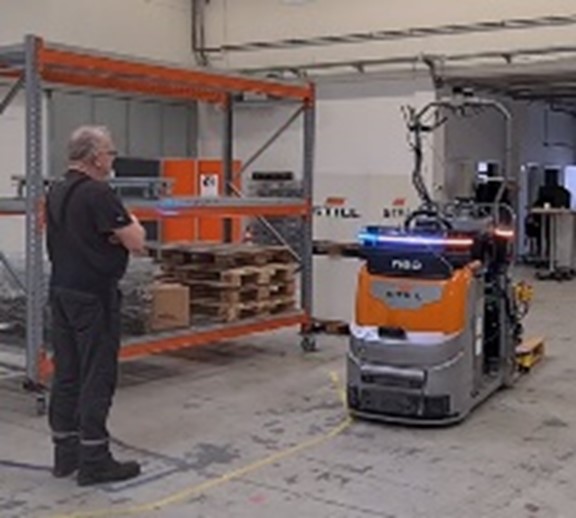}%
        \label{fig:rl_left}}
    \hfill
    \subfloat[][VR replication]{%
        \includegraphics[width=0.48\columnwidth]{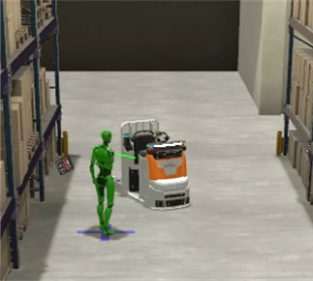}%
        \label{fig:vr_left}}
    
    \caption{Participants interacting with an AGV: (a) real-world encounter and (b) corresponding replication in the VR environment.} 
    \label{fig:RealWorldvsVR}
\end{figure}

\begin{figure*}[!t]
    \vspace{0.2cm}
    \centering
    \includegraphics[width=2\columnwidth]{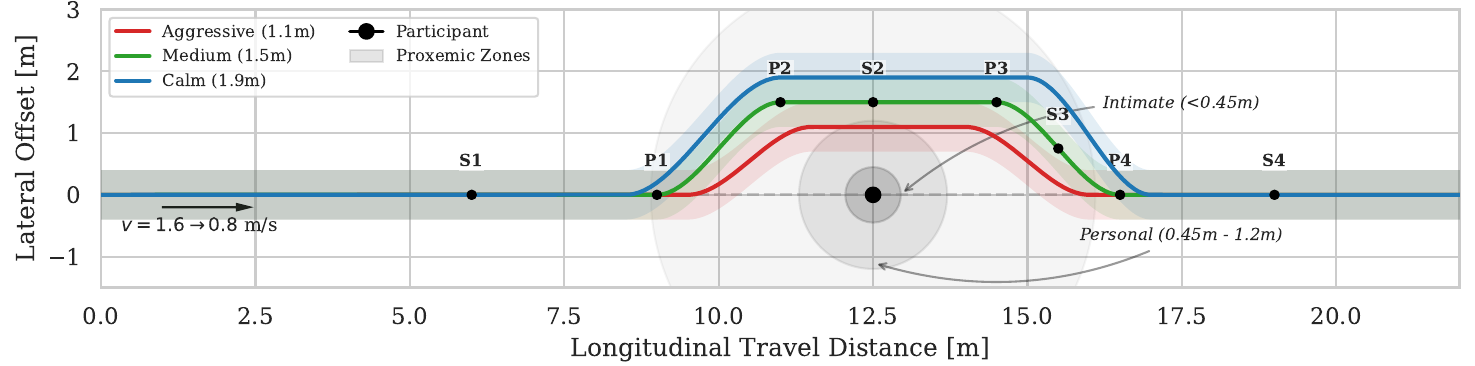}
    \caption{Three AGV trajectory profiles (aggressive, medium and calm) around the static participant, positioned at 12.5 m. Circles around the participant show the three proxemics zones \cite{hall1968proxemics}.}
    \label{fig:agv_trajs}
\end{figure*}

In practice, evaluating the user experience of a moving robot is challenging, as it must be set up in the target environment and all potential risks posed by a large, heavy robot must be eliminated. Testing in a simulation using virtual reality (VR) or augmented Reality (AR) would significantly reduce effort and eliminate the risk of collisions with test subjects \cite{wiczorek2022evaluation}. The question of whether findings from the VR/AR testing can be transferred to the real-world settings is currently being investigated in numerous research projects \cite{maruhn2021vr},
%[HA4.1][AH4.2]
e.g. in the automotive industry and for delivery robots. Schneider and Bengler \cite{schneider2020virtually} state that a quantitative transfer of outcomes from VR to reality should be approached cautiously due to moderate correlations across many studies and the limited number of studies explicitly aimed at evaluating comparability.

%\subsection*{Problem Statement and central question}
Furthermore, existing studies of perceived safety in encounters with AGVs focus either on small service robots that do not pose a threat to humans or on autonomous vehicles (AV). In contrast, the robots in intralogistics are much larger and more dangerous than service robots, and travel at much lower speeds than AVs. Additionally, many academic studies recruit university students as test participants, who have a rather different background compared to the industry workers. Finally, most studies do not employ continuous threat assessment measures reported during the encounter, but instead rely on post-hoc questionnaires.

In this paper, we examine perceived safety during encounters with an industrial AGV among real factory workers. We compare three trajectory profiles, which differ in the distance to the person. In addition to the questionnaires, we use the pressure-sensitive controller trigger as a continuous measure of perceived threat, and compare the real-world experience to the VR replication. Finally, we ask participants to set their preferred collision avoidance parameters in VR. Our study aims to address the following research questions:
\begin{itemize}
    \item How does the perceived safety in encounters with industrial AGVs relate to the passing distance?
    \item To which extent the perceived safety in encounters with industrial AGVs in VR compares to a real-world experience?
    %\item Can the desired driving parameters determined in the virtual environment be used to design suitable movement behavior in a real environment? What are the limitations or necessary adjustments? 
    \item What are the preferred collision avoidance parameters (the lateral and longitudinal distances) from the perspective of a standing person?
    % Do we want to add this topic, or better leave it out as before? I just thought this aspect might also not be in scope for the workshop, since it is about human motion, not suitable robot motion. I added some findings in the discussion section about this as well.
\end{itemize}

% RQ1: Are worker perceptions of an industrial AGV's driving behavior in VR comparable to those in a real-world encounter?

% RQ2: Can driving parameters tested in VR be transferred to a real AGV deployment, and what are the corrections?

\section{Related work}

A worker's safety in shared spaces with industrial-level AGVs is a critical issue that ISO standards \cite{ISO3691-4} aim to address. However, these norms do not address perceived safety and threat levels people experience in proximity to these platforms, which are often expressed as hesitation or reduced walking speeds \cite{haney2026perceived}. Perceived safety is shown to depend on the approach direction of the robot, its speed, and appearance \cite{haney2024literature}. However, the majority of the experiments still take place in laboratory settings and do not leverage commercially deployable AGVs \cite{haney2024literature}. Furthermore, most studies do not employ continuous measures reported during the encounter but instead rely on post-hoc questionnaires \cite{haney2024literature}, which often miss the moment when perceived danger spikes, which could inform the collision avoidance behavior and intent communication design for AGVs.

Exposing participants to large heavy vehicles in real-world encounters can lead to injuries and is also logistically challenging, especially in real production environments. Therefore, in domains such as autonomous driving \cite{brill2023external} and industrial AGV interaction \cite{han2025interactions}, Virtual Reality (VR) is a well-established method. VR has been used to evaluate the effect of explicit communication from AGVs (with and without an LED eHMI) on trust and perceived safety \cite{han2025interactions}. A central question for this line of research, though, is the transferability of VR to the real world. For HRI applications, researchers found that people accept closer distances to physical robots than to virtual ones \cite{li19comparing}, that trust and engagement transfer well for service robots \cite{plomin2023virtual} and for industrial manipulators \cite{legler2023emotional}. However, some researchers point to task dependency as a factor in transferability \cite{mielke2025virtual}, while others question it altogether \cite{esterwood2025virtually}. To the best of our knowledge, there are no systematic studies of perceived safety transferability between VR and the real factory floor for large AGVs. In this work, we adopt measures such as continuous trigger press \cite{de2019external, bazilinskyy2021should}, which is tightly coupled to proximity to the AGV, and apply them to large-payload AGVs across matched VR and real-world conditions. 

% \begin{itemize}
%     \item The use of physical interfaces, e.g. buttons, for perceived threat level evaluation in user studies
%     \item Safety assessment in AD using VR
% \end{itemize}

\section{Methods}

\subsection{Study Design}

% To compare the UX of an encounter with a real autonomous mobile robot (AMR) and a virtual representation, the same traffic situation was built in both environments. Since the real MR and the virtual representation needed to behave in the same reproducible way, a simple traffic scenario for a non-moving person was developed. The maneuver always had to be conducted in the exact same way, thus for the test the MR was pre-programmed.

% % Real world MR
% Within the research project IMOCO4.E, the order picker truck iGo neo OPX from Still GmbH was transferred into an MR. The navigation system of the modified truck is based on the ROS2 Navigation stack. Since all planners in the navigation stack generate an autonomous reactive behaviour in the test a cubic-spline-planner developed within IMCOC4.E was used in combination with the Navigation2-RPP-plugin (RPP = Regulated Pure Pursuit) with a vehicle-adapted calibration parameter set. During the user tests, the demonstrator truck displayed optical and acoustical signals according to the external HMI (eHMI) concept described in the VDMA recommendation \cite{scherb2023design}. Since the base truck was not capable of displaying multicolored visual signals or extended acoustic signals, a prototypical setup was implemented. Four RGB-LED strips and a broadband speaker were added to the truck.

In this paper, we investigate the perceived safety of people in interactions with AGVs, experienced in virtual reality and in the real world. For this purpose, we design a collision avoidance scenario in an intralogistic environment in which the robot follows a pre-programmed maneuver around the standing participant (see Fig.~\ref{fig:RealWorldvsVR}). The robot platform is an iGo neo OPX order picker (Still GmbH), modified into an autonomous research vehicle as part of the IMOCO4.E project\footnote{\url{https://www.imoco4e.eu}}. Navigation is managed via the ROS2 stack, utilizing a custom cubic-spline planner coupled with the Regulated Pure Pursuit (RPP) plugin for precise, vehicle-specific path tracking. To facilitate human-robot interaction, the platform is augmented with a prototypical external HMI (eHMI) system comprising four RGB-LED strips and a broadband speaker, supporting the optical and acoustical signals aligned with the VDMA recommendation \cite{scherb2023design}.

Trajectory of the robot is parametrized by a sequence of control points that define its evasive behavior relative to a stationary participant (see Fig. \ref{fig:agv_trajs}). The robot approaches the participant along a linear path at a max. velocity of $1.6\,\text{m/s}$, then decelerates to $0.8\,\text{m/s}$ until $S_1$, where an amber flashing signal is initiated via the eHMI to communicate intent. The evasive maneuver begins at the turn-in point $P_1$, which is adjustable along the longitudinal axis to vary the onset of the swerve. The lateral displacement and the duration of the parallel bypass are governed by points $P_2$ and $P_3$, while $P_4$ determines the longitudinal coordinate for the return to the original path. Acceleration back to the initial velocity commences at point $S_4$. Utilizing this geometric framework, we define three discrete trajectory profiles: an \textit{aggressive} version with minimal permissible safety distances, a \textit{calm} version with generous clearance, and a \textit{medium} version serving as an intermediate baseline. The distances in Fig.~\ref{fig:agv_trajs} are specified relative to the person's position and the center point of the robot's footprint.

% \begin{figure}[!t]
% \vspace{0.2cm}
%      \centering
%      \subfloat[][Worker interacting with AGV]{\includegraphics[width=4cm]{figures/rl_left.jpg}\label{fig:rl_left}}
%      \subfloat[][AGV closeup]{\includegraphics[width=4cm, height=3.6cm]{figures/rl_right.jpg}\label{fig:rl_right}}
%      \subfloat[][Worker interacting with AGV]{\includegraphics[width=4cm]{figures/vr_left.png}\label{fig:vr_left}}
%      \subfloat[][AGV closeup]{\includegraphics[width=4cm, height=3.6cm]{figures/vr_right.jpg}\label{fig:vr_right}}
%      \caption{VR-scenario}    
%      \label{fig:RealWorldvsVR}
% \end{figure}

% VR representation
The virtual reality part of the experiment is created in Unity. The robot behavior and signaling are comparable across the real-world and VR versions of the study. To ensure a realistic experience with sufficient immersion, we use a shaded CAD model and implement realistic driving sounds. The simulation runs on a stationary PC using SteamVR and streams wirelessly to a VR headset (HTC XR Elite). %The choice of headset is mainly determined by the availability of a business edition without a log-in and a steady internet connection, which is important for use inside a company, and, secondly, by the capability for reliable broadband wireless streaming using WiFi 6. Because test persons had to be able to move within an area of 10x3m, a large movement area was important, which is possible with markerless inside-out tracking. 

In addition to experiencing the predefined profiles, we asked the participants to parameterize their custom trajectories using a VR interface. The interface utilizes virtual representations of the AGV at each control point, which participants can manipulate via a VR controller to adjust the path's geometry. The resulting trajectory is visualized as a projection on the virtual floor. Participants can conduct virtual test-runs and are permitted unlimited iterations to refine their design.
%Due to the kinematic constraints of the physical platform, specifically, the safety-rated deceleration triggered by high-yaw-rate maneuvers.
The self-defined trajectories are executed exclusively in the virtual environment to prevent artificial restrictions on the participants' preferred safety boundaries.

\subsection{Data Collection and Metrics}

The study was conducted at the STILL GmbH in Hamburg, Germany, a major manufacturer of forklift trucks and other material-handling equipment. The location has approximately 3000 employees. All test subjects were recruited from the local workforce. %It is still a member of the KION Group AG, with over 40,000 employees worldwide. 
The study cohort ($N=10$) comprised professional staff from industrial assembly and internal logistics divisions, balanced for age (50\% aged 20--49, 50\% aged $\geq 50$) and technical experience, with 40\% of participants reporting significant prior AGV interaction. The gender distribution included 9 males and 1 female participant, all of whom were active employees in truck assembly or forklift operation at Still GmbH.

We employ a multi-modal approach to quantify user perception, combining continuous real-time input with discrete psychometric scales. During each trial, participants provided a continuous measure of perceived threat level $T(s) \in [0,1]$ via a pressure-sensitive trigger on the VR controller, recorded with respect to the total travel distance of the robot $s$. Controller input ranges from a null value (fully relaxed) to a maximum value (maximum perceived risk). From $T(s)$ we derive three per-trial metrics: the \emph{trigger use rate},
the percentage of trials in which the trigger was engaged at
any point; the \emph{total duration} ($t_{\text{total}}$), the
cumulative time the trigger was held active; and the
\emph{area under the curve} (AUC),
\begin{equation}
\text{AUC} = \int_{s_{\text{start}}}^{s_{\text{end}}} T(s)\,\mathrm{d}s,
\label{eq:auc}
\end{equation}
with $s_{\text{start}}$ and $s_{\text{end}}$ bounding the
analysis window along the AGV path.

Following each run, a discrete assessment is recorded using a 5-point Likert scale ranging from \textit{``much too unsafe''} to \textit{``much safer than necessary''}. To evaluate the effect of modality (VR and real-world),
%the direct comparison phase is followed by a targeted questionnaire assessing perceptual differences between the physical and virtual environments.
we asked the participants to rate the similarity in
(a) relative safety ($1=\text{much safer in real}, 5=\text{much safer in VR}$),
(b) velocity perception ($1=\text{much faster in VR}, 5=\text{much faster in real}$),
(c) spatial proximity ($1=\text{much closer in VR}, 5=\text{much closer in real}$), and
(d) overall encounter perception ($1=\text{very strong}, 5=\text{not at all}$). %Furthermore, we measured psychological immersion in the virtual environment using the iGroup Presence Questionnaire (IPQ) \cite{schwind2019using}, administered immediately after the VR blocks. 
These quantitative metrics are augmented by semi-structured qualitative interviews conducted at the midpoint and the end of the study to record insights into participants' perception of the ``reality gap.''

\begin{figure*}[!t]
    \centering
    \includegraphics[width=1.9\columnwidth]{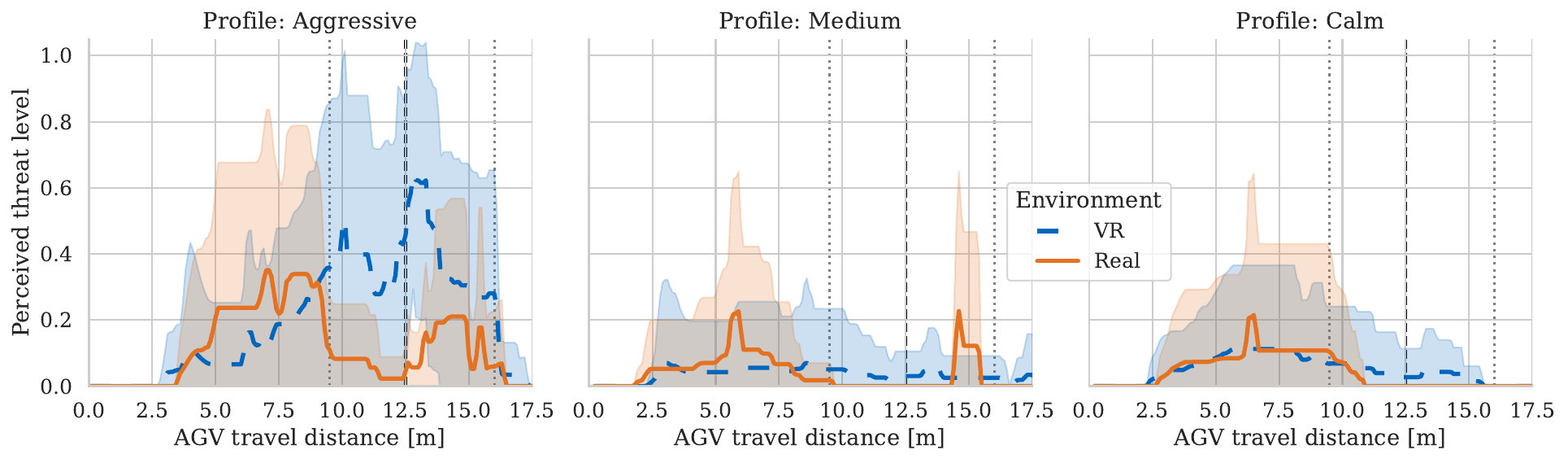}
    \caption{Evolution of the perceived threat levels with proximity to the AGV along the three trajectory profiles, recorded using continuous trigger press. Vertical lines denote the key maneuvers (P1, S2 and P4 from Fig.~\ref{fig:agv_trajs}).}
    \label{fig:perceived_safety_traj}    
\end{figure*}

\section{Results}

Perceived threat levels based on the trigger press are summarized in Fig.~\ref{fig:perceived_safety_traj}. Both in VR and real-life, participants pressed the trigger most actively in the
\textit{aggressive} profile; the \textit{medium} and \textit{calm}
trajectories elicited only low-amplitude signals. %The trigger-press duration was substantially longer in VR than in the real world, with greater inter-participant variance in VR. 
%Aggregating the perceived threat level traces against robot travel distance reveals that the \textit{aggressive} response shape also differed qualitatively between modalities. In the real world it was bimodal: participants pressed the trigger early during the approach (before the turn-in $P_1$), released it as the robot passed them, and pressed again as the robot returned to the straight path. In VR the response was unimodal, beginning at $P_1$, intensifying through the bypass, and receding steeply once the robot accelerated again at $S_4$. The \textit{medium} profile in VR showed a low, near-constant signal with a small drop after the robot passed the participant; the real-world counterpart was lower still, with two narrow peaks at $S_1$ (end of deceleration) and near 15~m (end of the bypass). The \textit{calm} profile in VR resembled the \textit{medium} one in shape but with marginally higher magnitude, a small peak at $S_1$, and the trigger held continuously until the maneuver ended. In the real-world \textit{calm} condition, the trace tracked the VR profile up to $P_1$ but featured a more pronounced $S_1$ peak and no further response once the maneuver had begun.
The trigger presses in the real encounter were more pronounced at the beginning of the collision avoidance maneuver, whereas in the VR the trigger reactions were registered through the trial.

Table~\ref{tab:threat_metrics_compact} aggregates the
three objective metrics across profiles and modalities.
All three decreased from \textit{aggressive} to \textit{calm}
in both modalities, consistent with reduced perceived threat
as the lateral clearance grew. The cross-modality offset was
most pronounced in the \textit{aggressive} profile, where
VR produced a 75\,\% higher mean AUC ($3.5 \pm 3.3$ vs.\
$2.0 \pm 2.0$), a 67\,\% longer mean $t_{\text{total}}$
($6.4 \pm 4.8$~s vs.\ $3.8 \pm 3.7$~s), and a
15-percentage-point higher trigger use rate (90\,\% vs.\
75\,\%). $t_{\text{total}}$ remained higher in VR for both the
\textit{medium} (1.8 vs.\ 1.1~s) and \textit{calm}
(2.4 vs.\ 0.9~s) profiles, with the largest \emph{relative}
VR/RW gap appearing in the \textit{calm} profile. The two
non-aggressive profiles otherwise showed small and
inconsistent differences: AUC differed by 0.1 in
\textit{medium} and 0.3 in \textit{calm}, and the
trigger use rate favoured VR for \textit{aggressive} and
\textit{calm} but the real world for \textit{medium}
(50\,\% vs.\ 40\,\%). Inter-participant variance was large
due to a small sample size.

\begin{figure}[!t]
    % \centering
    % \includegraphics[width=\columnwidth]{figures/Fig3_Revised.png}
    % \caption{Aggregated total duration of perceived risk for the aggressive trajectory profile across Virtual Reality (VR) and Real-World (RW) modalities.}
    % \label{fig:duration_unsafe}
    % \centering
    % \includegraphics[width=\columnwidth]{figures/rating_safety_plot_separate.png}
    % \caption{Safety ratings (5-point Likert scale) for three robot trajectory profiles across Virtual Reality (VR) and Real-World environments.}
    % \label{fig:subj_safety}
    % \centering
    % \includegraphics[width=\columnwidth]{figures/cross_modality_plot_horizontal_blended.png}
    % \caption{Subjective participant perception (VR vs. Real World) on a 5-point Likert Scale: (a) Overall encounter similarity, (b) Relative distance, (c) Relative speed, and (d) Relative safety.}
    % \label{fig:cross_modal}
    \centering
    \raisebox{-17pt}{\includegraphics[width=0.51\columnwidth]{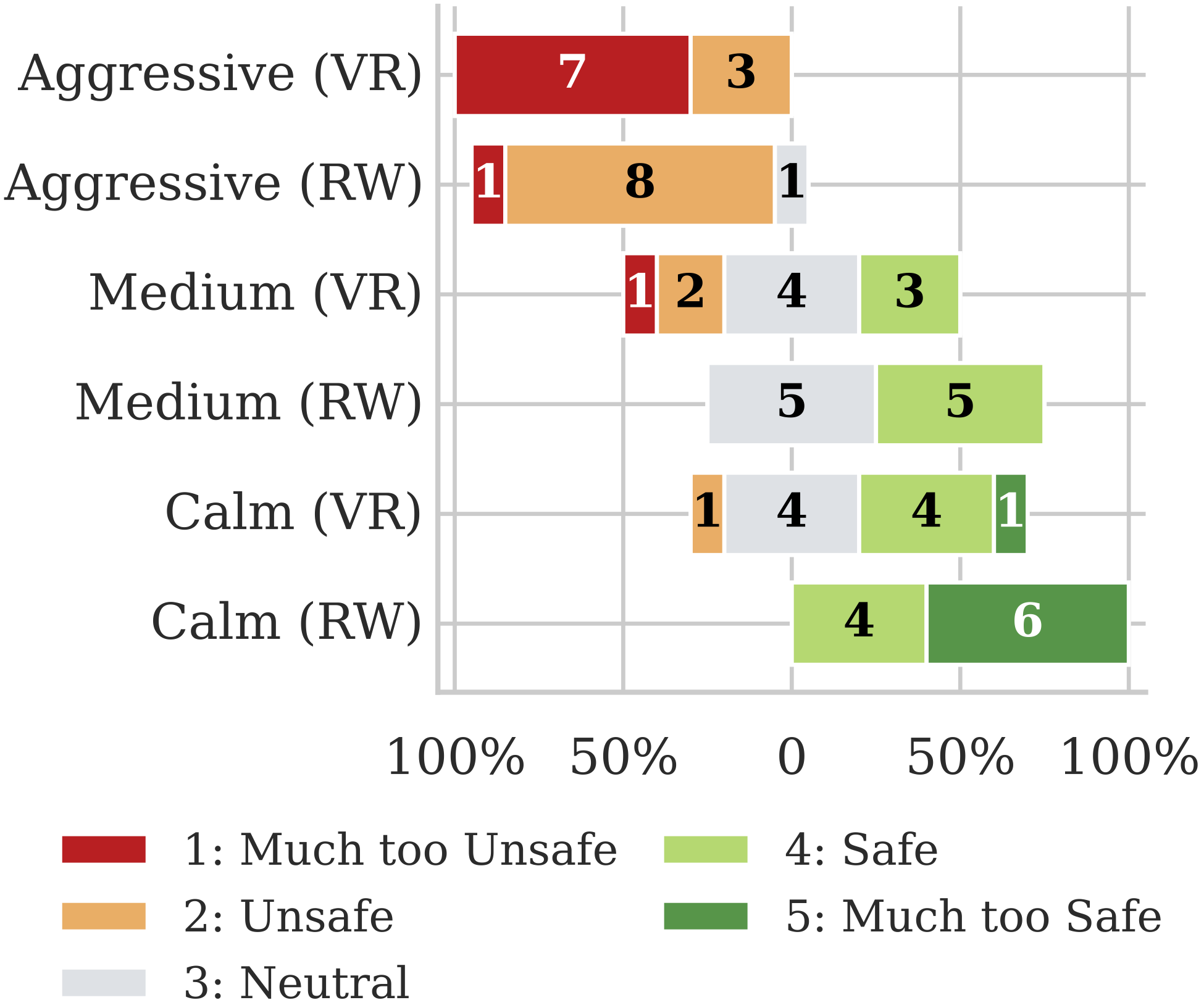}}
    \includegraphics[width=0.47\columnwidth]{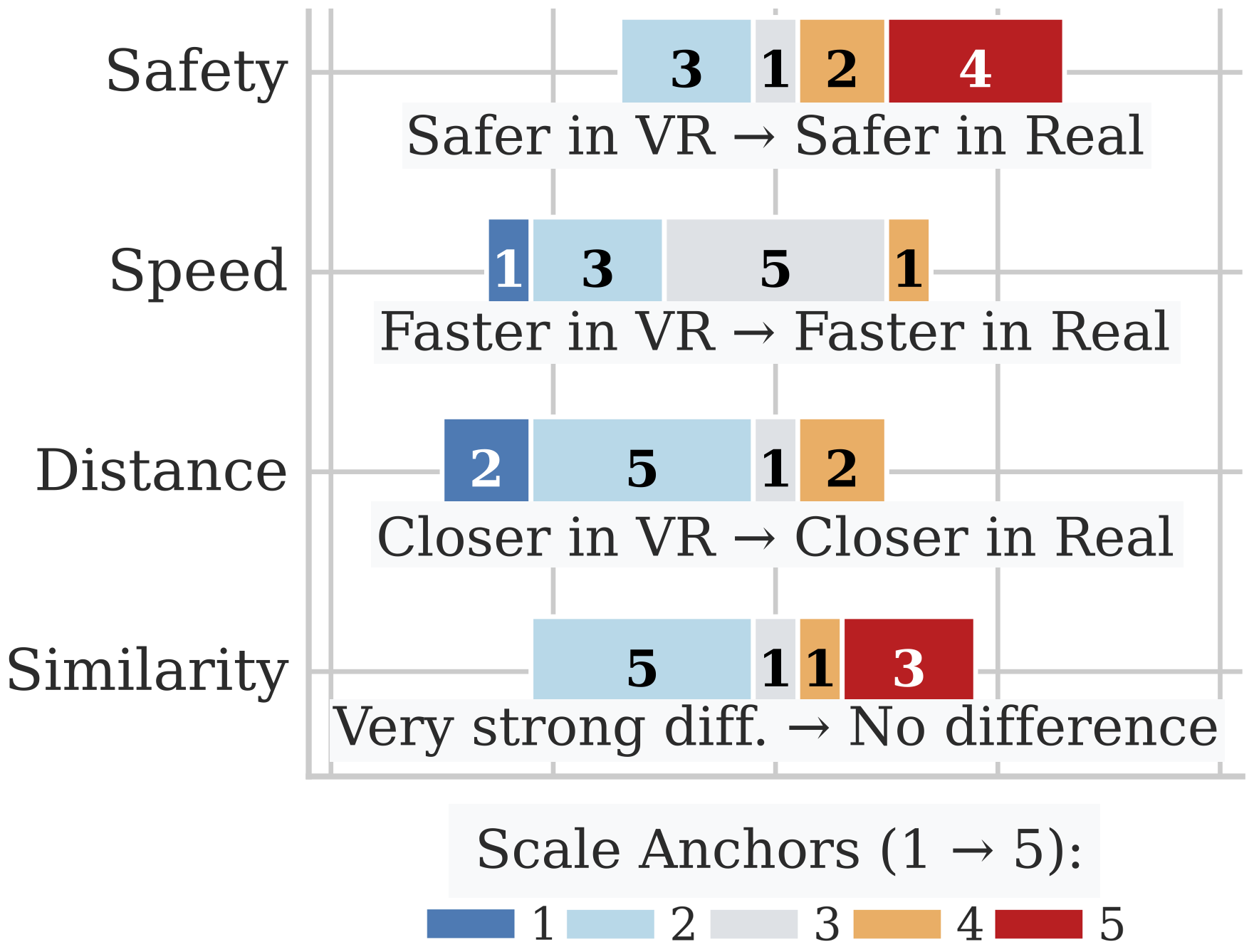}
    \caption{Subjective Measures (5-pt Likert scale). \textbf{Left:} Safety ratings for three robot trajectory profiles across Virtual Reality (VR) and Real-World (RW) modalities (M.t. = Much too). \textbf{Right:} Cross-modality perception differences: (a) relative safety, (b) relative speed, (c) relative distance, and (d) overall encounter similarity.}
    \label{fig:subjective_results}
\end{figure}

Subjective safety ratings increased across profiles from
\textit{aggressive} to \textit{calm}
(Fig.~\ref{fig:subjective_results} left). In the \textit{aggressive}
profile, both modalities clustered at the unsafe end:
VR ratings spanned 1--2, while most real-world encounters
were rated 2. For the \textit{medium} profile, medians were
comparable but the spread differed in direction --- VR
extended downward to 1, real-world upward to 5. The largest
median offset appeared in the \textit{calm} profile, where
the VR median (3, spread 2--4) lay one point below the
real-world median (4, spread 4--5). Averaged across the
three predefined profiles, VR ratings were 0.7 points below
their real-world equivalents.

From the VR and real-world similarity
questionnaire (Fig.~\ref{fig:subjective_results} right),
(a) relative safety ranged from 1 to 4 with no ratings at 5 (mean 2.3), consistent with the directional bias seen in Fig.~\ref{fig:subjective_results}; (b) relative speed and (c) relative distance had narrow spreads concentrated between 2 and 3 (means 2.6 and 2.3), with consistent reports that distances felt closer and speeds higher in VR; and
(a) overall encounter similarity had a wide spread from 2 to 5 (mean 3.2).

\begin{table}[t]
\centering
\footnotesize % Reduced font size to fit in one column
\caption{Perceived threat metrics across the three trajectory profiles and two modalities (Mean $\pm$ SD)}
\label{tab:threat_metrics_compact}
\begin{tabular}{@{}llcccc@{}} % @{} removes outer padding
\toprule
\textbf{Profile} & \textbf{Modality} & \textbf{Trigger rate [\%]} & \textbf{AUC} & \textbf{$t_{\text{total}}$ [s]} \\
\midrule
\multirow{2}{*}{Aggressive} & Real & 75 & $2.0 \pm 2.0$ & $3.8 \pm 3.7$ \\
                            & VR & 90 & $3.5 \pm 3.3$ & $6.4 \pm 4.8$ \\
\addlinespace
\multirow{2}{*}{Medium}     & Real & 50 & $0.7 \pm 1.3$ & $1.1 \pm 2.3$ \\
                            & VR & 40 & $0.6 \pm 1.7$ & $1.8 \pm 3.4$ \\
\addlinespace
\multirow{2}{*}{Calm}       & Real & 20 & $0.7 \pm 2.0$ & $0.9 \pm 2.5$ \\
                            & VR & 30 & $0.9 \pm 2.0$ & $2.4 \pm 4.7$ \\
\bottomrule
\end{tabular}
\label{tab:threat_metrics_compact}
\end{table}

% \begin{figure}[!t]
%     \centering

% \end{figure}

% \begin{figure}[!t]

% \end{figure}

\section{Discussion}

The perceived threat level measurements are consistent with the Likert-scale ratings after each test run. Both show that participants feel slightly less safe in VR. This result is consistent with the questionnaire results after the direct comparison, where participants, on average, felt slightly less safe in the VR environment, judged distances slightly closer, and reported a slightly higher speed. This interpretation is also consistent with the answers during the qualitative interview: seven participants stated that the results would be well transferable, one said there would be some differences, and two said comparison would be difficult. %Additionally, it was stated that a final test should be done using a real vehicle. 
Interestingly, three participants stated that they felt less safe in VR, even though, in the back of their minds, they always knew that, being in a virtual space, they were in no real danger. %Overall, we conclude that the experiences regarding perceived safety, perceived distance, and speed are generally comparable between VR and real-life.
%, thus answering research question one with a yes. Regarding research question 2, there are differences in experiences with distance and speed estimation, as well as perceived safety.

%Regarding the IPQ results, a comparison with \cite{maruhn2021vr} was made. He used the IPQ to evaluate the effect of an avatar on immersion. The result in that study, averaged across the setups, was: SP: 4,9; INV: 3,2; REAL: 3,1; G: 4,7. A comparison of his results with ours shows high values across all subscales, except Experienced realism, which is due to differences in image quality. In the experiment of \cite{maruhn2021vr}, photorealistic environments were used. Given that values for spatial presence and involvement are quite high, and the interviews indicated that the test participants felt very involved and present in the VR environment, the quality of the VR application can be considered sufficient.

\begin{figure}[t]
    \centering
    \includegraphics[width=\columnwidth]{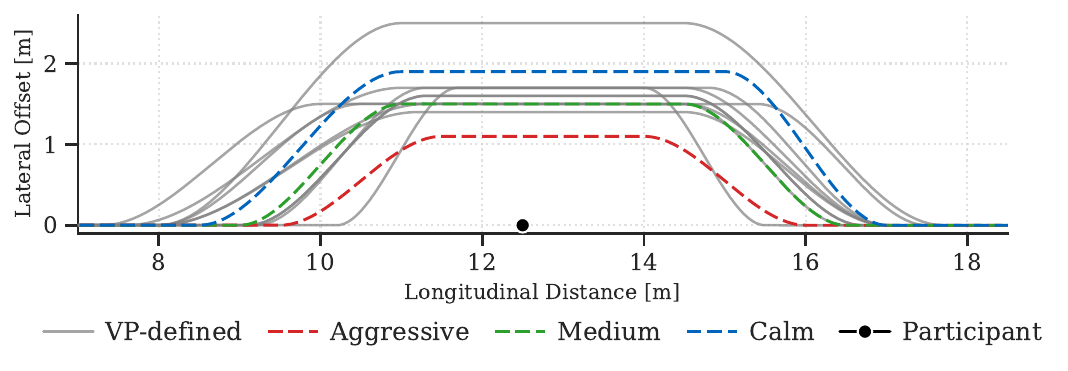}
    \caption{Preferred collision avoidance profiles of participants in VR.}
    \label{fig:self-trajectories}
\end{figure}

Regarding the desired parameters of the collision avoidance behavior, participants reported the lateral distance and the start of the collision avoidance maneuver as the main sources of discomfort. The self-defined trajectories, shown in Fig.~\ref{fig:self-trajectories}, reflect that a combination of the medium profile (in terms of the lateral offset) and the calm profile (in terms of the longitudinal distance) is preferred. %The medium distance defined was 1.3 m. Since the test participants stated distances would seem closer in VR compared to real-world and given the ratings of the real-world medium version a distance of 1,1 m from the center of the participant to the edge of the AMR seems appropriate. The second characteristic often mentioned in the interviews was the distance to the participant when the maneuver would start. In the self-defined trajectory, an average of 2,3 m was defined (see Figure 116). Given again the difference in perception of distances, the good ratings of the medium real-world version and the comments in the interviews the distance of 1,8 m seems appropriate in this case. 

\paragraph*{Limitations}
With $N{=}10$, our findings with respect to the perceived safety and preferred distance are indicative.
%we cannot statistically quantify differences in perceived distance or speed. These limitations are primarily due to the use of a real ${>}1000$~\si{\kilo\gram} industrial truck from an active deployment site and to the fact that only active factory workers could participate, all of whom had prior exposure to comparable AGVs. Our exploration of RQ3 is therefore only qualitative, 
Determining the safe parameters for collision avoidance would require a larger sample size to achieve statistical significance, and should consider the robot's speed as a factor in perceived safety. Furthermore, our findings are based on a single scenario with a stationary participant. Further encounter types need to be considered in future work \cite{brayan2025navwareset}. Finally, while the gender distribution among participants mirrors the actual workforce at the production site, future studies should include more balanced distributions to avoid potential biases.
%Real and virtual trucks differed in two visible respects: a structural arch and side-mounted safety scanners are present only on the real unit, potentially biasing perceived mass, illusion~\cite{schmidtler2016size}. VR signal lights were dimmer than their physical counterparts, but participants did not notice this in the post-hoc interviews. Insights into driving parameters are derived from a single scenario with a stationary participant, with blinker onset ${\sim}2$~\si{\sec} prior to evasion and completed deceleration to it.

%\section{Conclusion}

%Given the results of this study, that UX evaluations of MR driving behavior in VR are generally comparable to real-world tests, the next step is to evaluate autonomous driving behavior in VR. For this task, an enhanced VR setup was created that enables test persons to enter a robotic simulation at runtime. The ROS2-based simulation, including a ROS2 navigation stack, is connected to a VR headset capable of rendering the required visualization. Both applications continuously exchange positional data of all relevant assets in the scene. The test person can see the MRs in the virtual environment, and their position is sent back to the simulation, enabling the simulated MRs to respond to their movements. The aim is to implement human-aware navigation behavior for the MRs and conduct UX tests to evaluate participants' impressions.

% \begin{figure}
%     \centering
%     \includegraphics[width=\columnwidth]{ROS_Unity bridge.jpg}
%     \caption{ROS-VR bridge}
%     \label{fig:enter-label}
% \end{figure}

\newpage
\bibliographystyle{IEEEtran}
\bibliography{bibliothek}

\end{document}